\documentclass[12pt]{article}
\evensidemargin\oddsidemargin

\usepackage{latexsym,url}
\usepackage{graphicx}

\usepackage{amssymb}
\usepackage{epstopdf}
\usepackage{amsmath}
\usepackage{amsthm}

\newtheorem{thm}{Theorem}
\newtheorem{cor}[thm]{Corollary}
\newtheorem{fact}[thm]{Fact}
\newtheorem{lem}[thm]{Lemma}

\newcommand{\ctr}[1]{\begin{center} #1 \end{center}}

\newcommand{\NN}{\mathbf{N}}

\newcommand{\Prob}{{\rm Prob}}

\newcommand{\myproof}{\noindent {\em Proof.}  }

\newcommand{\QED}{\hfill $\qed$}
\newcommand{\dom}{\mbox{\rm dom}}
\newcommand{\bin}{\mbox{\rm  bin}}

\newcommand{\fdom}{\mbox{\rm \footnotesize{dom}}}
\newcommand{\fbin}{\mbox{\rm \footnotesize{bin}}}

\if01
{
\noindent {\bf #1 \hfill #2} \\[-2ex]
\begin{flushright}\begin{minipage}{\textwidth}
   }%
{\end{minipage}
\end{flushright}}
\fi

\title{\bf 
Most 
Programs Stop Quickly or Never Halt}

\author{Cristian S. Calude$^{1}$
and Michael A. Stay$^{2}$\\
\phantom{xx}\\
$^{1}$Department of Computer Science\\ University of
Auckland, New Zealand\\ Email: {\tt
cristian@cs.auckland.ac.nz}\\
\phantom{xx}\\
$^{2}$Department of Mathematics\\
University of California Riverside, USA\\
Email: {\tt
mike@math.ucr.edu}}
   
\begin{document}
\date{}

\maketitle 
\begin{abstract} 
The aim of this paper is to provide a probabilistic, but non-quantum, analysis of the Halting Problem.  Our approach is to have the probability space extend over both space and time and to consider the probability that a random $N$-bit program has halted by a random time. We postulate {\it an a priori computable probability distribution on all possible runtimes} and we prove that given an integer $k>0$, we can effectively compute a time bound $T$ such that the probability that an $N$-bit program will eventually halt given that it has not halted by $T$ is smaller than $2^{-k}$.  

We also show that the set of halting programs (which is computably enumerable, but not computable) can be written as a disjoint union of a computable set and a set of effectively vanishing probability.

Finally, we show that ``long'' runtimes are effectively rare. More formally, the set of times at which an $N$-bit program can stop after the time $2^{N +\,\mbox{constant }}$  has effectively zero density.

\end{abstract}

\thispagestyle{empty}

\section{Introduction}

The Halting Problem for Turing machines is to decide whether an arbitrary Turing machine $M$ eventually halts on an arbitrary input $x$. As a Turing machine $M$ can be coded by a finite string---say, $code(M)$---one can ask whether there is  a Turing machine $M_{\footnotesize halt}$ which, given  $code(M)$ and the input $x$, eventually stops and produces $1$ if $M(x)$ stops,  and $0$ if $M(x)$ does not stop. Turing's famous result states that this problem cannot be solved by any Turing machine, i.e.~there is no such  $M_{\footnotesize halt}$. Halting computations can be recognised by simply running them; the main difficulty is to
detect non-halting programs. In what follows time is discrete.

Since many real-world problems arising in the fields of compiler optimisation, automatised software engineering, formal proof systems, and so forth are equivalent to the Halting Problem, there is a strong interest---not merely academic!---in understanding the problem better and in providing alternative solutions.

There are two approaches we can take to calculating the probability that an $N$-bit program will halt.  The first approach, initiated by Chaitin \cite{chaitin75},  is to have the probability space range only over programs; this is the approach taken when computing the Omega number, \cite{CDS,cris}.  In that case, the probability is    $Prob_N = \#\{p \in \Sigma^N \mid p  \mbox{ halts}\}/2^N.$  For a self-delimiting machine, $Prob_N$ goes to zero when $N$ tends to infinity, since it becomes more and more likely that any given $N$-bit string is an extension of a shorter halting program.  For a universal non-self-delimiting Turing machine, the probability is always nonzero for large enough $N$: after some point, the  universal non-self-delimiting Turing machine will simulate a total Turing machine (one that halts on all inputs), so some fixed proportion of the space will always contribute.  The probability in this case is uncomputable, machine-dependent; in general, 1 is the best computable upper bound one can find. In this approach it matters only whether a program halts or not; the time at which a halting program stops is irrelevant.

Our approach is to have the probability space extend over both space and time, and to consider the probability that a random $N$-bit program---which hasn't stopped by some given time---will halt by a random later time.  In this approach, the stopping time of a halting program is paramount. The problem is that there is no uniform distribution on the integers, so we must choose some kind of distribution on times as well.  Any distribution we choose must have that most long times are rare, so in the limit, which distribution we choose doesn't matter very much.

The new approach was proposed by Calude and Pavlov \cite{CP} (see also \cite{ACP}) where a mathematical quantum ``device" was constructed to probabilistically solve the Halting Problem.  The procedure has two steps: a) based on the length of the program and an  {\it a priori} admissible error $2^{-k}$, a finite time $T$ is effectively computed, b) a quantum ``device", designed to work on a randomly chosen test-vector is run for the time $T$; if the device produces a click, then the program halts; otherwise, the program probably does not halt, with probability higher than $1-2^{-k}$. This result uses an unconventional model of quantum computing, an infinite dimensional Hilbert space.  This  quantum proposal has been discussed in  Ziegler \cite{Z}.

 It is natural to ask whether the quantum mechanics machinery is essential for obtaining the result. In this paper we discuss a method to ``de-quantize'' the algorithm.  We have been motivated by some recent approximate solutions to the Halting Problem obtained by K\"ohler,  Schindelhauer  and  M. Ziegler \cite{KSZ} and experimental work \cite{CDS,LP}.\footnote{For example, Langdon and Poli \cite{LP} suggest that, for a specific universal machine that they describe, about $N^{-1/2}$ programs of length $N$ halt.}
Different approaches were proposed by Hamkins and Miasnikov 
\cite{HM}, and D'Abramo \cite{D'A}.

Our working hypothesis, crucial for this approach, is to postulate {\it an a priori computable probability distribution on all possible runtimes}. Consequently, the probability space is the product of the space of programs of fixed length (or of all possible lengths), where programs are uniformly distributed, and the time space, which is discrete and has an {\it a priori} computable probability distribution. In this context we show that given an integer $k>0$, we can effectively compute a time bound $T$ such that  the probability on the product space that an $N$-bit program will eventually halt given that it not stopped by $T$ is smaller than
$2^{-k}$. This phenomenon is similar to the one described for proofs in formal axiomatic systems \cite{CJ}.

We also show that for every integer $k >0$, the set of halting programs (which is computably enumerable, but not computable) can be written as  a disjoint union of a computable set and a set of probability effectively smaller than  $2^{-k}$.

Of course, an important question is to what extent the postulated hypothesis is acceptable/realistic. The next part of the paper deals with this question offering an argument in favor of the hypothesis.  First we note that for any computable probability distribution most long times are effectively rare, so in the limit they all have the same behavior irrespective of the choice of the distribution. Our second argument is based on the common properties of the times when programs may stop. Our proof consists of three parts: a) the exact time a program stops is algorithmically not too complicated; it is (algorithmically) nonrandom because most  programs either stop `quickly' or never halt, b) an $N$-bit program which hasn't stopped by time $2^{N +\, \mbox{constant }}$ cannot halt at a later random time, c) since nonrandom times are (effectively) rare, the density of times an $N$-bit program can stop vanishes.

We will start by examining a particular case in detail, the behavior of all programs of length 3 for a certain Turing machine.  This case study will describe various possibilities of halting/non-halting programs, the difference between a program stopping exactly {\em at} a time and a program stopping {\em by} some time, as well as the corresponding probabilities for each such event.

Finally, we show some differences between the halting probabilities for different types of universal machines.

Some comments will be made about the ``practicality'' of the results presented in this paper: can we use them to approximately solve
any mathematical problems?

\section{A case study}

We consider all programs of length $N=3$ for a simple Turing machine $M$  whose domain is the finite set $\{000,010,011,100,110,111\}$. The halting ``history'' of these programs, presented in Table~1, shows the times at which the programs in the domain of $M$ halt. The program $M(p_{1})$ halts at time $t=1$, so it is indicated by an ``h'' on the row corresponding to $p_{1}$ and time $t=1$; the program $M(p_{4})$ hasn't halted at time $t=5$, so on the row corresponding to $p_{4}$ and time $t=4$ there is a blank. Finally, programs which haven't stopped by time $t=17$ never stop.

\begin{center}
\begin{tabular}{|c|c|c|c|c|c|c|c|c|c|}\hline
Program/time & $t=1$ & $t=2$ & $t=5$ & $t=8$ & $\ldots$ & $ t=14$ & $t=15$ & $t=16$ & $t=17$\\
\hline
$p_{1}=000$ & h & h & h & h & h & h & h & h & h \\
\hline
$p_{2}=001$ &   &   &   &   &   &   &   &   &  \\
\hline
$p_{3}=010$ &   &   &   &   &   &   & h & h & h \\
\hline
$p_{4}=011$ &   &   &   & h & h & h & h & h & h \\
\hline                              
$p_{5}=100$ &   &   &   &   &   & h & h & h & h \\
\hline
$p_{6}=101$ &   &   &   &   &   &   &   &   &  \\
\hline
$p_{7}=110$ & h & h & h & h & h & h & h & h & h \\
\hline
$p_{8}=111$ &   &   &   &   &   &   &   & h & h\\
\hline\end{tabular}\\

\bigskip

Table~1:  Halting  ``history'' for 3-bit programs.
\end{center}

Here are a few miscellaneous facts we can derive from Table~1:
\begin{itemize}
\item the program  $M(p_{1})$ halts exactly at time $t=1$,
\item the set of 3-bit programs which halt exactly at time $t=1$ consists of $\{p_{1}, p_{7}\}$, so the `probability' that a randomly chosen 3-bit program halts at time $t=1$ is 
$\#\{\mbox{3-bit programs halting at time 1}\}/ \#\{\mbox{3-bit programs}\} = \#\{p1, p7\}/8 = 2/8=1/4$,

\item the set of 3-bit programs which halt  by time $t=8$ consists of $\{p_{1},  p_{4}, p_{7}\}$, so the `probability' that a randomly picked  3-bit program halts by time $t=8$ is 
$\#\{$3-bit programs halting by time $8\}/ \#\{\mbox{3-bit programs}\} = \#\{p1, p4, p7\}/8 = 3/8$,

\item the `probability' that a random 3-bit program eventually stops is $\#\{$3-bit programs that halt$\}/ \#\{$3-bit programs$\}=6/8$,

\item the program  $M(p_{4})$ hasn't stopped by time $t=5$, but stops at time $t=8$,

\item the `probability' that a 3-bit program does not stop by time $t=5$ and that it eventually halts is $\#\{$3-bit programs that eventually halt that have not stopped by time $t=5\}/\#\{\mbox{3-bit programs}\}  = \#\{p3, p4, p5, p8\}/8 = 4/8 = 1/2$,

\item the `probability' that a 3-bit program eventually stops given that it has not halted by time $t=5$ is $\#\{$3-bit programs that eventually halt that have not stopped by time $t=5\}/\#\{$3-bit programs that have not halted by time  $t=5\} = 4/(8-2) = 2/3,$

\item the `probability' that a 3-bit program halts at time $t=8$ given that it has not halted by time $t=5$ is $\#\{$3-bit programs that halt by time $t=8 \mbox{ but not by time } t=5\}/\#\{$3-bit programs that have not halted by time $ t=5\} = 1/6$.
\end{itemize}

We can express the above facts in a bit more formal way as follows.  We fix a universal Turing machine $U$ (see section~3 for a definition) 
and a pair $(N,T)$, where $N$ represents the length of the program and $T$ is the ``time-window'', i.e.\ the interval $\{1,2,\ldots ,T\}$, where the computational time is observed.  The probability space is thus \[Space_{N,T} = \Sigma^{N} \times \{1,2,\ldots ,T\}.\]

Both programs and times are assumed to be {\it uniformly distributed}. For $A \subseteq Space_{N,T}$ we define  $\Prob_{N,T}(A)$ to be $\#(A)\cdot 2^{-N} \cdot T^{-1}$.

\medskip

Define
\[A_{N,T} = \{(p,t)\in Space_{N,T}  \mid U(p) \mbox{ stops exactly at time $t$}\},\]
\noindent and
\[B_{N,T} = \{(p,t)\in Space_{N,T}  \mid U(p) \mbox{ stops  by time $t$}\}.\]

\medskip

\begin{fact}
\label{prel}  We have: $\Prob_{N,T}(A_{N,T})  \le  \frac{1}{T}$ and  $\Prob_{N,T}(B_{N,T}) \le  1.$ 
\end{fact}

\myproof It is easy to see that $\# (A) \le 2^{N}$, consequently,
$$\Prob_{N,T}(A_{N,T}) = \frac{\#(A_{N,T})}{2^{N} \cdot T} \le \frac{2^{N}}{2^{N} \cdot T} = \frac{1}{T}, \phantom{x} \Prob_{N,T}(B_{N,T}) \le \frac{\#(B_{N,T})}{2^{N} \cdot T}\le  \frac{2^{N} \cdot T}{2^{N} \cdot T} = 1.$$   \QED

\medskip

\noindent {\bf Comment}. The inequality  $\Prob_{N,T}(B_{N,T}) \le  1$ does not seem to be very informative. However, for all $N$, one can construct a universal Turing machine $U_N$ such that $\Prob_{N,T}(B) = 1;$  $U_{N}$ cannot be self-delimiting (see, for a definition,  section~4).  There is no universal Turing machine  $U$ such that $\Prob_{N,T}(B) = 1, $ for all $N$, so  can we do better than stated in Fact~\ref{prel}?

More precisely, we are interested in the following problem: 
\begin{quote} \it
We are given a universal Turing machine $U$ and a randomly chosen program $p$ of length $N$ that we know  not to stop by time $t$.
Can we effectively evaluate the `probability' that $U(p)$ eventually stops? 
\end{quote}

An obvious way to proceed is the following. Simply, run in parallel all programs of length $N$ till the time  $T_{N} = \max\{t_{p} \mid  |p|=N, U(p) \mbox{  halts} \} = \min\{t\mid \mbox{for all } |p|=N, t_{p} \le t\}$, where $t_{p} $ is the exact time $U(p)$ halts (if indeed it stops). In other words, get the analogue of the Table~1 for $U$ and $N$, and then calculate directly all probabilities. This method, as simple as it may look, isn't very useful, since the time $T_{N}$ is {\it not computable}   because of the undecidability of the Halting Problem.

Can we overcome this serious difficulty?

\if01
\medskip

If $U$ is a self-delimiting universal Turing machine, then

\[\Omega_{U} = \sum_{N\ge 1} \#\{ p \mid |p|=N, U(p) \mbox{  halts}\}\cdot 2^{-N},\]

\noindent hence

\[\lim_{N \rightarrow \infty}   \#\{ p \mid |p|=N, U(p) \mbox{  halts}\}\cdot 2^{-N}=0.\]

I suspect we can prove that this limit is {\it not constructive}.

\medskip

Claim 4:   If $U$ is a self-delimiting universal Turing machine, then $$\lim_{N \rightarrow \infty}\Prob_{N,T}(B) \le \lim_{N \rightarrow \infty}
\frac{ \#\{ p \mid |p|=N, U(p) \mbox{  halts}\}\ \cdot T}{2^{N} \cdot T} =0.$$

Claim 4 doesn't talk about asymptotic
behavior in time, but rather in program length.  I.e. it says,  ``long
programs run a long time on average".
\fi

\section{Notation}

All strings are binary and the set of strings is denoted by $\Sigma^{*}$.  The length of the string $x$ is denoted by $|x|$. The logarithms are binary too.  Let $\NN = \{1,2,\ldots\}$ and let $\bin:\NN \rightarrow \Sigma^*$ be the computable bijection which associates to every $n\ge 1$ its binary expansion without the leading 1,  
\ctr{\begin{tabular}{r|c|c|c}
$n$&$n_2$&$\bin(n)$& $|\bin(n)|$\\
\hline
1&1&$\lambda$ & 0\\
2&10&0 & 1\\
3&11&1& 1\\
4&100&00 & 2\\
$\vdots$&$\vdots$&$\vdots$&$\vdots$
\end{tabular}}

We will work with  Turing machines $M$  which process  strings into  strings. The domain of $M$, $\dom (M)$, is the set of strings on which $M$ halts (is defined).  The  {\it natural complexity} \cite{cs1} of the string $x\in\Sigma^*$ (with respect to $M$) is $\nabla_M(x) = \min\{n\ge 1 \mid M({\rm bin}(n))=x\}$.  The Invariance Theorem \cite{cris} has the following form: we can effectively construct a machine $U$ (called {\em universal}) such that for every  machine $M$, there is a constant $\varepsilon>0$ (depending on $U$ and $M$) such that $\nabla_U (x) \leq \varepsilon \cdot \nabla_M (x)$, for all strings $x$. For example, if $U(0^{i}1x) = M_{i}(x)$ (where $(M_{i})$ is an effective enumeration of all Turing machines), then $\nabla_{U}(x) \le (2^{i+1}+1)\cdot \nabla_{M_{i}}(x)$, because $0^{i}1\bin(m) = \bin (2^{i+1 +\lfloor \log (m)\rfloor}+m)$, for all $m\ge 1$. In what follows we will fix a universal Turing machine $U$ and we will write $\nabla$ instead of $\nabla_{U}$. There are some advantages in working with the complexity $\nabla$ instead of the classical complexity $K$ (see \cite{cris});  for example, for every $N>0$, the inequality $\#\{x\in \Sigma^{*}\,:\, \nabla (x) <N\}\le N$ is  obvious; a better example appears in \cite{cs0} where $\nabla$ is a more natural measure to investigate the relation between incompleteness and uncertainty.
 

\section{Halting according to a computable time distribution}

\if01 
We start by observing the physical context for the scenario investigated here.  The running of $U(p)$ is supposed to be done in some physical universe which determines the ``local runtime''; the observer, possibly in another physical universe, has its own ``time'' and ``time probability distribution''.  For example, $U$ may run on a satellite revolving around the Earth and the observer could be located somewhere in New Zealand.\footnote{There may or may not be a clear relation between the machine runtime and observer's time. }
\fi

We postulate {\it an a priori computable probability distribution on all possible runtimes}. Consequently, the probability space is the product of the space of programs---either taken to be all programs of a fixed length, where programs are uniformly distributed, or to be all programs of all possible lengths, where the distribution depends on the length---and the time space, which is discrete and has an {\it a priori} computable probability distribution. 

In what follows we randomly choose a time $i$  from according to a  probability distribution $\rho(i)$ which effectively converges to 1, that is, there exists a computable function $B$ such that for every $n\ge B(k)$, $$|1-\sum_{i=1}^{n}\rho(i)|< 2^{-k}.$$  

How long does it take for an $N$-bit program $p$ to run without halting on $U$ to conclude that the probability that $U(p)$ eventually stops is less than $2^{-k}$?
 
It is not difficult to see that the probability that an $N$-bit program which hasn't stopped on $U$ by time $t_{k}$ (which can be effectively computed) will eventually halt is not larger than $\sum_{i\ge t_{k}} \rho(i)$, which effectively converges to 0, that is, there is a computable function $b(k)$ such that for $n\ge b(k)$, $\sum_{i\ge n} \rho(i) < 2^{-k}$.
   
The probability distribution $\rho(i)$ may or may not be related to the computational runtime of a program for $U$. Here is an example of a probability distribution which effectively converges  to 1 and relates the observer time to the computational runtime. This probability distribution is reminiscent of Chaitin's halting probability \cite{cris}, but in contrast, is {\it computable}.

The idea is to define the distribution at moment $i$ to be $2^{-i}$ divided by the exact time it takes $U(\bin(i))$ to halt, or to be 0 if $U(\bin(i))$ does not halt.  Recall that $t_{p}$ is the exact time $U(p)$ halts (or $t_{p}=\infty$ when $U(p)$ does not halt). 

First we define the number
\[\Upsilon_{U}=\sum_{i\ge 1}2^{-i}/t_{\fbin(i)}. \]
It is clear that $0<\Upsilon_{U}<1$. Moreover,  $\Upsilon_{U}$ is computable.  Indeed, we construct an algorithm computing, for every positive integer $n$, the $n$th digit of $\Upsilon_{U}.$ The proof is simple: only the terms $2^{-i}/t_{\fbin(i)}$ for which $U(\bin(i))$ does not halt, i.e.\ $t_{\bin(i)}=\infty$,  produce `false' information because at every finite step of the computation they appear to be non-zero when, in fact, they are zero! The solution is to run all non-stopping programs $U(\bin(i))$ for enough time such that their cumulative contribution is too small to affect the $n$th digit of $\Upsilon_{U}$: indeed, if $n>2$, and $t_{\bin(i)}= 1$, for $i\ge n$, then $\sum_{i=n}^{\infty} 2_{-i}/t_{\fbin(i)}< 2^{-n}$.

So,  $\Upsilon_{U}$  induces a natural probability distribution on the runtime: to $i$ we associate\footnote{Of course, instead of $2^{-i}/t_{\fbin(i)}$ we can take $r_i/t_{\fbin(i)}$, where $\sum_{i\ge 1}r_i <\infty$, effectively.}
\begin{equation}
\label{timed}
\rho(i)= \frac{2^{-i}}{t_{\fbin(i)}\cdot \Upsilon_{U}}\raisebox{0.5ex}.
\end{equation}

The probability space is 
\[Space_{N, \{\rho(i)\}} = \Sigma^{N} \times \{1,2,\ldots \},\]
\noindent where $N$-bit programs are assumed to be uniformly distributed, and we choose at random a runtime from distribution (\ref{timed}).

\medskip

\begin{thm}
Assume that $U(p)$ has not stopped by time $T> k - \lfloor \log \Upsilon_{U} \rfloor$. Then, the probability (according to the distribution (\ref{timed})) that $U(p)$ eventually halts is smaller than $2^{-k}$.
\end{thm}

\myproof It is seen that 
\[     \frac{1}{\Upsilon_{U}} \sum_{i=T}^{\infty} \frac{2^{-i}}{t_{\fbin(i)}} 
   \le \frac{1}{\Upsilon_{U} \cdot 2^{T-1}} < 2^{-k},\]
for $T> k - \lfloor \log \Upsilon_{U} \rfloor$. The bound is computable because $\Upsilon_{U}$ is computable.

\QED

We now consider the probability space to be 
\[Space_{\{\rho(i)\}} = \Sigma^{*} \times \{1,2,\ldots \},\]
\noindent where $N$-bit programs  are assumed to be uniformly distributed, and the runtime is chosen at random from the computable probability distribution  $\{\rho(i)\}$. 

\medskip

\begin{thm}
Assume that $U$ and $Space_{\{\rho(i)\}}$ have been fixed.  For every integer $k >0$, the set of halting programs for $U$ can be written as  a disjoint union of a computable set and a set of probability effectively smaller than  $2^{-k}$.
\end{thm}

\myproof Let $b$ be a computable function such that for $n\ge b(k)$ we have  $\sum_{i\ge n} \rho(i) < 2^{-k}$.  The set of halting programs for $U$ can be written as  a disjoint union of the computable set $\{(p,t_{p}) \mid t_{p} < 2^{b(k+|p|+2)}\}$ and the set $\{(p,t_{p}) \mid 2^{b(k+|p|+2)} \le t_{p}  < \infty \}$. The last set has probability effectively less than 
\[      \sum_{N=1}^{\infty} \sum_{n=b(k+N+2)}^{\infty} \rho_{n}
    \le \sum_{N=1}^{\infty} 2^{-N-k-2} = 2^{-k-1}.\]
\QED

\medskip

\noindent {\bf Comment}. A stronger (in the sense that the computable set is even polynomially decidable), but machine-dependent, decomposition theorem for the set of halting programs was proved in \cite{HM}.

\section{How long does it take for a halting program to stop?}

The common wisdom says that it is possible to write short programs which stop after a very long time. However, it is less obvious that there are only a few such programs; these programs are ``exceptions''. 

Working with self-delimiting Turing machines, Chaitin \cite{busy} has given the following estimation of the complexity\footnote{Chaitin used  program-size complexity.} of the runtime of a program which eventually halts: there is a constant $c$ such that if $U(\bin(i))$ halts in time $t$, then 
\begin{equation}
\label{chaitin-ineqR}\nabla (\bin (t)) \le 2^{|\fbin(i)|}\cdot c \le i\cdot c.
\end{equation} 
Here $t$ is the first time $U(\bin(i))$ halts.\footnote{Of course, if $U(\bin(i))$ halts in time $t$, it stops also in time $t'>t$, but only finitely many $t'$  satisfy the inequality (\ref{chaitin-ineqR}). For the reader more familiar with the program-size complexity $H$---with repsect to a universal self-delimiting Turing machine \cite{cris}---the inequality (\ref{chaitin-ineqR}) corresponds to $H(\bin(t)) \le |\bin(i)| +c$.}   The above relation puts a limit on the {\it complexity}
of the time $t$ 
 a program $\bin(i)$, that eventually halts on $U$, has to run before it stops; this translates into a  limit on the  time  $t$
because only finitely many strings have complexity bounded by a constant.  In view of (\ref{chaitin-ineqR}), {\it the bound depends only upon the  length of the program}; the program itself (i.e. $\bin(i)$) does not matter.

Because $\lim_{t\rightarrow \infty}\nabla (\bin(t)) = \infty$, there are only finitely many integers $t$ satisfying the inequality (\ref{chaitin-ineqR}). That is, there exists a critical value $T_{\footnotesize critical}$ (depending upon $U$ and  $|\bin(i)|$) such that if for each $t < T_{\footnotesize critical}$, $U(\bin(i))$ does not stop in time $t$, then $U(\bin(i))$ never halts. In other words, 
\begin{quote}
if $U(\bin(i))$ does not stop in time $T_{\footnotesize critical}$, then $U(\bin(i))$ never halts. 
\end{quote}

So, what prevents us from running the computation $U(\bin(i))$ for the time $T_{\footnotesize critical}$ and deciding whether it halts or not?  Obviously, the uncomputability of $T_{\footnotesize critical}$.  Neither the natural complexity $\nabla$ nor any other size complexity, like $K$ or $H$, is computable (see \cite{cris}).  Obviously, there are large integers $t$ with small complexity $\nabla(\bin(t))$, but they cannot be effectively ``separated'' because  we cannot effectively compute a bound $b(k)$ such that $\nabla (\bin(t))>k$ whenever $t > b(k)$.

The above analysis suggests that {\it a program that has not stopped after running for a long time has smaller and smaller chances to eventually stop}. The bound (\ref{chaitin-ineqR}) is not computable.  Still, can we ``extract information'' from the inequality (\ref{chaitin-ineqR}) to derive a computable probabilistic description of this phenomenon?

Without loss of generality, we assume that the universal Turing machine $U$ has a built-in counting instruction. Based on this, there is an effective transformation which for each program $p$ produces a new program $time(p)$ such that there is a constant $c>0$ (depending upon $U$) for which the following three conditions are satisfied:
\begin{enumerate}
\item $U(p)$ halts iff $U(time(p))$ halts,
\item $|time(p)| \le |p| + c$,
\item if $U(p)$ halts, then it halts at the step $t_{p} = \bin^{-1}(U(time(p)))$.
\end{enumerate}

Intuitively, $time(p)$ either calculates the number of steps $t_{p}$ till $U(p)$ halts and prints $\bin(t_{p})$, or, if $U(p)$ is not defined, never halts. The constant $c$ can be taken to be less than or equal to 2, as the counting instruction is used only once, and we need one more instruction to print its value; however, we  don't need to print the value $U(p)$. 

\medskip

We continue with a proof of the inequality (\ref{chaitin-ineqR}) for an arbitrary universal Turing machine.
  
\medskip

\begin{thm}
\label{nablaineq}
Assume that $U(p)$ stops at time $t_{p}$, exactly. Then,
\begin{equation}
\label{chaitin-ineqRR}\nabla (\bin (t_{p})) \le 2^{|p|+c+1}.
\end{equation}
\end{thm}
\myproof First we note that for every program $p$ of length at most $N$, $\bin^{-1}(p) < 2^{N+1}$. Indeed, $|p| =|\bin(\bin^{-1}(p))| \le N$ implies 
\begin{equation}
\label{expineq}
2^{|p|} \le \bin^{-1} (p) < 2^{|p|+1} \le  2^{N+1}.
\end{equation}

Since $U(p) = U(\bin(\bin^{-1}(p)))$ we have:
\[\nabla (U(p))  = \min\{i\ge 1\;:\; U(\bin(i))=U(p)\}\le \bin^{-1}(p),\]
\noindent hence
\[\nabla (\bin(t_{p}))  = \nabla (U(time(p)))  \le \bin^{-1}(time(p)) < 2^{|p| + c +1},\]
\noindent 
because $|time(p)| \le |p|+c$ and (\ref{expineq}). 
\QED

\section{Can a program stop at an algorithmically random time?}

In this section we prove that no program of length $N\ge 2$ which has not stopped by time $2^{2N+2c+1}$ will stop at an algorithmically random time.  Consequently, since  algorithmically nonrandom times are (effectively) rare, there are only a few times an $N$-bit program can stop in a suitably large range.  As a consequence, the set of times at which an $N$-bit program can stop after the time $2^{N +\,\mbox{constant }}$ has effectively zero density.

\if01
In this scenario, an $N$-bit program which has not halted by time
$2^{2N+2c+1}$ has effectively vanishing to zero chances to stop a later time
(called exponential stopping time); the set of exponential stopping has effectively zero density.
\fi

\medskip 

A binary string $x$ is ``algorithmically random'' if $\nabla (x) \ge  2^{|x|}/|x|$. Most binary strings of a given length $n$ are  algorithmically random because they have high density:  $\#\{ x \in \Sigma^{*} \,:\, |x|=n, \nabla (x) \ge 2^{n}/n\}\cdot 2^{-n} \ge 1 - 1/n$ which tends to 1 when $n \rightarrow \infty$.\footnote{In the language of program-size complexity, $x$ is ``algorithmically random'' if $H(x) \ge |x| - \log (|x|)$.}

A time $t$ will be called ``algorithmically random'' if $\bin (t)$ is algorithmically random. 

\begin{thm}
\label{randstop}
Assume that an $N$-bit program $p$ has not stopped on $U$  by time $2^{2N+2c+1}$, where $N\ge 2$ and $c$ comes from Theorem~\ref{nablaineq}.  Then, $U(p)$ cannot exactly stop at any algorithmically random time $t \ge 2^{2N+2c+1}$.
\end{thm}

\myproof 
First we prove that for every $n\ge 4$  and $t\ge 2^{2n-1}$, we have: 
\begin{equation}
\label{ubound}
2^{|\fbin(t)|} > 2^{n} \cdot |\bin(t)|.
\end{equation}


Indeed, the real function $f(x) = 2^{x}/x$ is strictly increasing for $x \ge 2$ and tends to infinity when $x \rightarrow \infty$.  Let $m=|\bin(t)|$.  As $2^{2n-1} \le t < 2^{m+1}$, it follows that $m \ge 2n-1$, hence  $2^{m}/m \ge 2^{2n-1}/(2n-1) \ge 2^{n}$.  The inequality  is true for every $|\bin (t)| \ge  2n-1$, that is, for every $t\ge 2^{2n-1}$.

Next we take  $n=N+c+1$ in (\ref{ubound}) and we  prove  that every algorithmically random time $t \ge 2^{2N+2c+1}$, $N \ge 2$, does not satisfy the inequality (\ref{chaitin-ineqRR}). Consequently, no program of length $N$ which has not stopped by time $2^{2N+2c+1}$ will stop at an  algorithmically random time.

\QED

\medskip

A time $t$ is called ``exponential stopping time'' if there is a program $p$ which stops on $U$ exactly at    $t=t_{p} > 2^{2|p|+2c+1}$.  How large is the set of exponential stopping times?  To answer this question we first need  a technical result.

\medskip

\begin{lem}\label{series}
Let $m \ge 3, s\ge 1$. Then
\[\frac{1}{2^{s}-1}\cdot  \sum_{i=0}^{s} \frac{2^{i}}{m+i} < \frac{5}{m+s-1}\raisebox{0.5ex}.\]
\end{lem} 

\myproof
Let us denote by $x_{s}^{m}$ the left-hand side of the inequality below.  It is easy to see that $$ x_{s+1}^{m} = \frac{2^{s}-1}{2^{s+1}-1} \cdot x_{s}^{m} + \frac{2^{s+1}}{m+s+1}\le \frac{x_{s}^{m}}{2} + \frac{2}{m+s+1}
\raisebox{0.5ex}.$$

Next we prove by induction (on $s$) the inequality in the statement of the lemma. For $s=1$ we have $x_{1}^{m} = 1/m + 2/(m+1) < 5/m$. Assume that  $x_{s}^{m}< 5/(m+s-1).$ Then:
\[ x_{s+1}^{m} \le  \frac{x_{s}^{m}}{2} + \frac{2}{m+s+1} < \frac{5}{2(m+s-1)} + \frac{2}{m+s+1}\le \frac{5}{m+s}\raisebox{0.5ex}. \]
\QED

\medskip

The density of times in the set $\{1,2,\ldots ,N\}$ satisfying the property $P$ is the ratio $\#\{ t \mid
1 \le t \le N, P(t)\}/N$. A property $P$ of times has ``effective zero density'' if   the density of times satisfying the property $P$  effectively  
converges to zero, that is, 
there is a computable function $B(k)$ such that for every
$N> B(k)$, the density of times satisfying the property $P$ is smaller than $2^{-k}$.

\medskip

\begin{thm}
\label{onestep}
For every length $N$, we can effectively compute a threshold time $\theta_{N}$ (which depends on $U$ and $N$) such that if a program of length $N$ runs for time $\theta_N$  without halting, then the density of times greater than $\theta_N$ at which the program can stop has effective zero density. More precisely, if an $N$-bit program runs for time  $T >\max\{\theta_{N}, 2^{2+5\cdot 2^{k}}\}$, then the density of times at which the program can stop is less than $2^{-k}$.
\end{thm}

\myproof 
We choose the bound $\theta_{N} = 2^{2N+2c+1}+1$, where $c$ comes from (\ref{chaitin-ineqRR}).  Let $T> \theta_{N}$ and put $m = 2N+2c+1$, and $s = \lfloor \log (T+1)\rfloor -m$.  Then, using Theorem~\ref{randstop}  and Lemma~\ref{series}, we have:
\[ \frac{1}{T-2^{m} + 1} \cdot \#\left\{2^{m}\le t\le T \mid \nabla (\bin(t)) \ge \frac{2^{|\fbin (t)|}}{|\bin(t)|}\right\} \phantom{xxxxxxxxxxxxxxxxxxxxxxxx}\]\\[-4ex]
\begin{eqnarray*}
&\ge & \frac{1}{T-2^{m} + 1}\cdot  \sum_{i=0}^{s}\#\left\{2^{m+i}\le t\le 2^{m+i+1}-1\mid \nabla (\bin(t)) \ge \frac{2^{|\fbin (t)|}}{|\bin(t)|}\right\}  \\[1ex]
&= & \frac{1}{T-2^{m} + 1}\cdot \sum_{i=0}^{s}\#\left\{2^{m+i}\le t\le 2^{m+i+1}-1\mid \nabla (\bin(t)) \ge \frac{2^{m+i}}{m+i}\right\}  \\[1ex]
&\ge & \frac{1}{T-2^{m} + 1}\cdot \sum_{i=0}^{s} 2^{m+i} \left( 1 - \frac{1}{m+i}\right)\\
& \ge & 1 - \frac{1}{T-2^{m} + 1}\cdot \sum_{i=0}^{s}   \frac{2^{m+i}}{m+i}
\end{eqnarray*}
\begin{eqnarray*}
&\ge &  1 - \frac{1}{(2^{s} - 1)}\cdot \sum_{i=0}^{s}   \frac{2^{i}}{m+i}\\
&> &  1 - \frac{5}{m+s-1}\raisebox{0.5ex},
\end{eqnarray*}

\noindent consequently, the density of algorithmically random times effectively converges to 1:

\[  \lim_{T \rightarrow \infty} \frac{\#\{t\mid t > \theta_{N}, t\le T, t \not= t_{p}, \mbox{  for all } p \mbox{  with }  |p|=N\}}{T-2^{m} + 1} \phantom{xxxxxxxxxxxxxxxxxxxxxxxxxxxxxxxxxxxxxxxxxxxxxx} \]\\[-7ex]

\begin{eqnarray*}
&\ge & \lim_{T \rightarrow \infty} \frac{1}{T-2^{m} + 1} \cdot \#\left\{2^{m}\le t\le T \mid \nabla (\bin(t)) \ge \frac{2^{|\fbin (t)|}}{|\bin(t)|}\right\} = 1,
\end{eqnarray*}

\noindent so the density of times greater than $\theta_N$ at which an $N$-bit program can stop  effectively converges to zero. 
\QED

\medskip

The next result states that  ``almost any'' time is not an exponential stopping time.

\medskip

\begin{cor}
The set of exponential stopping times has effective zero density.
\end{cor}

\myproof It is seen that 
\[ \{t \mid t=t_{p}, \mbox{  for some } p \mbox{  with  } t > 2^{2|p|+2c+1}\}  \phantom{xxxxxxxxxxxxxxxxxxxxxxxx}\]\\[-7ex]
\begin{eqnarray*}
&\subseteq & \bigcup_{N\ge 1}\left\{ t \mid  t >  2^{2|p|+2c-1}, |p|=N, \, \nabla (\bin(t)) < \frac{2^{|\fbin (t)|}}{|\bin(t)|}\right\}\\[1ex]
&\subseteq & \left\{ t \mid  t >  2^{2c+1},  \, \nabla (\bin(t)) < \frac{2^{|\fbin (t)|}}{|\bin(t)|}\right\}\raisebox{0.5ex},
\end{eqnarray*}

\noindent which has effectively zero density in view of 
Theorem~\ref{onestep}.

\QED
   
\section{Halting probabilities for different universal machines}

In this section we show a significant difference between the halting probability of a program of a given length for a universal Turing machine and for a universal self-delimiting  Turing machine: in the first case the probability is always positive, while in the second case the probability tends to zero when the length tends to infinity.

\medskip

The probability that an arbitrary string of length $N$ belongs to $A \subset \Sigma^{*}$ is $\Prob_{N}(A) = \#(A\cap \Sigma^{N})\cdot 2^{-N}$, where $ \Sigma^{N}$ is the set of $N$-bit strings. 

If $U$ is a universal Turing machine, then the probability that an $N$-bit program $p$ halts on $U$ is  $\Prob_{N}(\dom (U))$. 

\medskip

\begin{fact} 
Let $U$ be a universal Turing machine. Then, $\lim_{N \rightarrow \infty} \Prob_{N}(\dom (U)) >0$. 
\end{fact}
\myproof 
We consider the universal Turing machine $U(0^{i}x) = T_{i}(x)$, described in section~3.   If $N$ is sufficiently large, then there exists an $i <N$ such that $T_{i}$ is a total function, i.e. $T_{i}$  is defined on each input, so  $\#(\dom (U) \cap \Sigma^{N}) \ge 2^{N-i-1}$. For all  such  $N$'s, $\Prob_{N}(\dom (U)) \ge 2^{-i-1}>0$.  The result extends to any universal Turing machine because of universality.

\QED

\medskip

A convergent machine $V$ is a Turing machine such that its $\zeta$ number is finite:
$$\zeta_{V} = \sum_{\fbin (n) \in \fdom (V)} 1/n <\infty,$$ see \cite{cs1}.
The $\Omega$ number of $V$ is $\Omega_{V} =\sum_{N=1}^{\infty} \Prob_{N}(\dom (V))$. Because $\zeta_{V} < \infty $ if and only if $\Omega_{V} < \infty$, see \cite{cs1}, we get:

\medskip

\begin{fact} 
Let $V$ be a convergent machine. Then, $\lim_{N \rightarrow \infty} \Prob_{N}(\dom (V)) =0$. 
\end{fact}

\medskip

Recall that a self-delimiting  Turing machine  $V$ is a machine with a prefix-free domain. For such a machine, $\Omega_{V} < 1$, hence we have:

\medskip

\begin{cor} 
Let $V$ be a universal self-delimiting Turing machine.  Then $\lim_{N \rightarrow \infty} \Prob_{N}(\dom (V)) =0$. 
\end{cor}

\medskip

The probability that an $N$-bit program never stops on a convergent Turing machine tends to one when $N$ tends to infinity; this is not the case for a universal Turing machine. 
      
\section{Final comments}

We studied the halting probability using a new approach, namely we considered the probability space extend over both space and time, and the probability that a random $N$-bit program will halt by a random later time given that it hasn't stopped by some threshhold time.  We postulated {\it an a priori computable probability distribution on all possible runtimes}.  Consequently, the probability space is the product of the space of programs---either taken to be all programs of a fixed length, where programs are uniformly distributed, or to be all programs of all possible lengths, where the distribution depends on the length---and the time space, which is discrete and has an a priori computable probability distribution.   We proved  that given an integer $k>0$, we can effectively compute  a time bound $T$ such that  the probability that an $N$-bit program will eventually halt, given that it has not stopped by time $T$, is smaller than $2^{-k}$.  

We also proved that  the set of halting programs (which is computably enumerable, but not computable) can be written as  a disjoint union of a computable set and a set of probability effectively smaller than any fixed bound.

Finally we showed that runtimes much longer than the lengths of their respective halting programs are (effectively) rare. More formally, the set of times at which an $N$-bit program can stop after the time $2^{N +\,\mbox{constant }}$  has effectively zero density.

Can we use this type of analysis for developing a probabilistic approach for proving theorems? 

The class of problems which can be treated in this way are the ``finitely refutable conjectures''.  A conjecture is finitely refutable if verifying a finite number of instances suffices to disprove it \cite{CJL}.  The method seems simple: we choose  a natural universal Turing machine $U$ and to each such conjecture $C$ we can effectively associate a program $\Pi_{C}$ such that $C$ is true iff $U(\Pi_{C})$ never halts.  Running $U(\Pi_{C})$ for a time longer than the threshold will produce a good evidence of the likelihood validity of the conjecture.  For example, it has been shown \cite{CCD} that for a natural $U$, the length of the program validating the Riemann Hypothesis is 7,780 bits, while for the Goldbach's Conjecture the length of the program is 3,484 bits.

Of course, the choice of the probability distribution on the runtime is paramount.  Further, there are at least two types of problems with this approach. 

First, the choice of the universal machine is essential.  Pick a universal $U$ and let $p$ be a program such that $U(p)$ never stops if and only if a fixed finitely refutable conjecture (say, the Riemann Hypothesis) is true.  Define $W$ such that $W(1)=U(p)$ (tests the conjecture), and $W(0x)=U(x)$.  The Turing machine $W$ is clearly universal, but working with $W$ ``artificially'' makes the threshold $\underline{\theta}$ very small.  Going in the opposite direction, we can write our simulator program in such a way that it takes a huge number of steps to simulate the machine---say Ackermann's function of the runtime given by the original machine.  Then the new runtime will be very long, while the program is very short.  Or we could choose very powerful instructions so that even a ridiculously long program on the original machine would have a very short runtime on the new one.  

The moral is that if we want to have some real idea about the probability that a conjecture has a counter-example, we should choose a simulator and program that are  ``honest":  they should not overcharge or undercharge for each time-step advancing the computation.  This phenomenon is very similar to the fact that the complexity of a single string cannot be independent of the universal machine; here, the probability of halting cannot be independent of the machine whose steps we are counting.

Secondly, the threshold $T$ will increase exponentially with the length of the program $\Pi_{C}$ (of course, the length depends upon the chosen $U$). For most interesting conjectures the length is greater than 100, so  it   is hopeless to imagine that these computations can be effectively carried out (see \cite{ml} for an analysis of the maximum speed of dynamical evolution). It is an open question whether another type of computation (possibly, quantum) can be used to speed-up the initial run of the program.

\section*{Acknowledgements}
We are grateful to  M. Zimand for comments and suggestions which significantly improved the paper.
We are indebted to E. Calude, G. Chaitin, M. Dinneen, M. Dumitrescu, N. Hay, and  K. Svozil for illuminating discussions on the topic of this paper.


\begin{thebibliography}{999}
\setlength\itemsep{0pt}
\bibitem{ACP}
V.~A. Adamyan, C.~S. Calude, B.~S. Pavlov. Transcending the limits of Turing computability, in T. Hida, K. Sait\^o, S. Si  (ed.). Quantum Information Complexity. Proceedings of Meijo Winter School 2003, World Scientific, Singapore, 2004, 119--137.

\bibitem{cris} C. S. Calude.  Information and Randomness: An Algorithmic Perspective,  2nd Edition, Revised and Extended, Springer-Verlag, Berlin,  2002.

\bibitem{CCD}
C. S. Calude, Elena Calude, M. J. Dinneen. A new measure of the difficulty  of  problems,  J. Mult.-Valued Logic Soft Comput. 12 (2006), 285--307.

\bibitem{CDS} C.~S. Calude, M.~J. Dinneen and C.-K. Shu. Computing a glimpse of randomness,  Experiment. Math. 11, 2 (2002), 369--378. 
 
 
\bibitem{CJ} C. S. Calude, H. J{\"u}rgensen.  Is complexity a source of incompleteness? Adv.  in Appl. Math. 35 (2005), 1--15.

\bibitem{CJL} C. S. Calude, H. J\"{u}rgensen, S. Legg. Solving finitely refutable mathematical problems, in C. S. Calude, G. P\u{a}un (eds.).  Finite Versus Infinite.  Contributions to an Eternal Dilemma, Springer-Verlag, London, 2000, 39--52.

\bibitem{CP} C.  S. Calude, B. Pavlov. Coins, quantum measurements, and Turing's barrier,  Quantum Inf. Process. 1, 1--2 (2002), 107--127.

\bibitem{cs0} C.~S. Calude, M.~A. Stay. From Heisenberg to G\"odel via Chaitin,  Internat. J. Theoret. Phys.  44, 7 (2005), 1053--1065.


\bibitem{cs1} C.~S. Calude, M.~A. Stay. Natural halting probabilities, partial randomness, and zeta functions,  Inform. and Comput.
204 (2006), 1718--1739.

\bibitem{chaitin75} G. J. Chaitin. A theory of program size formally identical to information theory,  J. ACM  22 (1975),  329--340. 

\bibitem{busy}
G. J. Chaitin.  Computing the busy beaver function, in T. M. Cover, B. Gopinath (eds.).  Open Problems in Communication and Computation, Springer-Verlag, Heidelberg, 1987, 108--112.

\bibitem{D'A} G. D'Abramo. Asymptotic behavior and halting probability of Turing Machines, math.HO/0512390, April 2006.



\bibitem{KSZ}  S. K\"ohler, C. Schindelhauer, M. Ziegler. On approximating real-world halting problems, in M. Li\'skiewicz, R. Reischuk (eds.). Proc. FCT 2005, Lectures Notes Comput. Sci. 3623, Springer, Heidelberg, 2005, 454--466.

\bibitem{LP} W.~B. Langdon, R. Poli. The halting probability in Von Neuamnn architectures, in P. Collet et a. (eds.). EuroGP 2006, Lectures Notes Comput. Sci. 3905, Springer, Heidelberg, 2006, 225--237.

\bibitem{HM} J.~D. Hamkins, A. Miasnikov. The halting problem is decidable on a set of asymptotic probability one, Notre Dame J. Formal Logic 47, 4 (2006), 515--524.

\bibitem{ml} N. Margolus, L. B. Levitin. The maximum speed of dynamical evolution, Phys. D 120 (1998), 188--195.


\bibitem{Z}  M. Ziegler. Does quantum mechanics allow for infinite parallelism?  Internat. J. Theoret. Phys.  44, 11 (2005), 2059--2071.

\end{thebibliography}
\end{document}